\documentclass[aps,prb,reprint,amssymb,amsmath,superscriptaddress,floatfix]{revtex4-1}
\usepackage{amsmath}
\usepackage{amsfonts}
\usepackage{times}
\usepackage{mathrsfs}
\usepackage{graphicx}
\usepackage{float}
\usepackage{listings}
\usepackage{upgreek}

\begin{document}
	
	\title{A wet etching method for few-layer black phosphorus with an atomic accuracy and compatibility with major lithography techniques}
	\author{Teren Liu}
	\affiliation{Department of Materials Engineering, University of British Columbia}
	\author{Tao Fang}
	\affiliation{Department of Materials Engineering, University of British Columbia}
	\affiliation{Department of Physics and Astronomy, University of British Columbia}
	\affiliation{Stewart Blusson Quantum Matter Institute, University of British Columbia}
	\author{Karen Kavanagh}
	\affiliation{Department of Physics, Simon Fraser University}
	
	\author{Guangrui (Maggie) Xia}
	\email{guangrui.xia@ubc.ca}
	\affiliation{Department of Materials Engineering, University of British Columbia}
	
	\begin{abstract}
		This paper reports a few-layer black phosphorus thickness pattern fabricated by a top-down nanofabrication approach. This was achieved by a new wet etching process that can etch selected regions of few-layer black phosphorus with an atomic layer accuracy. This method is deep-UV and e-beam lithography process compatible, and is free of oxygen and other common doping sources. It provides a feasible patterning approach for large-scale manufacturing of few-layer BP materials and devices.  
		%We shall also mention the function for few layer BP thickness variation device fabrication.
	\end{abstract}
	\maketitle
	\section{Introduction}
	2D materials, such as graphene, few-layer black phosphorus (BP) and transition metal dichalcogenides (TMDCs) materials, have drawn extensive interests for electronic and optoelectronic device applications, due to their unique properties such as high carrier mobility \cite{Qiao2014,Warschauer1963,Morita1986}, high on-off ratio, anisotropic proprieties \cite{Tran2014,Wang2015} and low defect concentrations. Moreover, many 2D materials, especially few-layer black phosphorus, has a tunable band gap based on the atomic layer number \cite{Cai2014,Tran2014}, which made band-gap-engineering-based quantum devices possible. Various theoretical studies showed that those devices have many appealing features, such as super-low power consumption, high on-off ratio, low subthreshold swing and tunable-photodetection wavelengths. \cite{Liu2015,Liu2016, Chang2015,Chen2017,Klinkert2018,Li2018,Cao2015,Meng2018,Cao2015a,Zeng2014,Li2018}  However,  those quantum devices were never validated by experiments. 
	
	It is mainly due to two issues.  One is that there have been no reports on large-scale synthesis of BP with good crystallinity, and the fabrication of few-layer BP crystals has been relying on the exfoliation from bulk BP crystals \cite{Li2014,Castellanos-Gomez2014}. This problem made mass production of BP crystals with stable size and thickness nearly impossible. Another problem is that precise control of the atomic layer numbers of BP, especially in selected areas, is still lacking.  Although some of the thinning and etching methods, such as scanning tunneling microscope (STM)-based nanopatterning \cite{Liu2016a}, liquid-phase-based thinning \cite{Hanlon2015,Yasaei2015,Lewis2017,Ambrosi2017, Yang2018}, femto-second laser based oxygen oxidation \cite{Robbins2017}, thermal thinning \cite{Fan2017,Lin2017,Luo2017}, and plasma thinning/etching \cite{Park2017} were reported in recent years, there were several key problems. First, those methods were mainly achieved on full samples instead of selected regions of a single sample. Secondly, the controllability of the thickness is still far from atomic layer accuracy. Thirdly, oxidation-based thinning/etching involves oxygen defects or surface dislocations, which would greatly affect the electronic properties and stability of BP after processing. Lastly, most of the thinning and etching methods mentioned require extra experiment conditions or equipment, which are not convenient. Common top-down approaches use photo-lithography and etching to achieve patterning and precise thickness control, which have not been reported for BP.
	
	In this work, using lithography and a new wet etching method  developed, we achieved few-layer BP patterns with a height difference of 20 nm (20 atomic layers) . This method has an atomic layer accuracy, and is compatible with deep ultra-violate (DUV) and electron-beam (e-beam) lithography processes. Concentration dependent etch rates from 0.5 to 2 nm/min along $<$010$>$ direction were observed. The samples were characterized by atomic force microscopy (AFM), Raman spectroscopy, energy-dispersive X-ray spectroscopy (EDS) and electron energy loss spectroscopy (EELS) method for the thickness, crystal structure, and element components. The crystallinity and the elements were intact after etching. AFM results confirmed that the etching recipe has a high anisotropy along  $<$010$>$ directions. 
	\section{Selection of a suitable etching reaction}
	
	Due to the lack of reports on chemical reactions with black phosphorus, we started with red phosphorus (RP) chemical reactions. RP and BP have very close standard enthalpy of formation and standard molar entropy. Common reactions with RP published from 1920's can be classified into five categories: 1) reactions with oxygen or high oxidative oxyacid \cite{Venugopalan1956,Yoza1965}, 2) reactions with high oxidative metal ions or some kind of metals, \cite{Rosenstein1920,Walker1926,BECKER1995,Han2000} 3) reactions with halogens \cite{Venugopalan1957,Largy1983,DORFMAN1994}, 4) strong Lewis base reactions \cite{Gusarova1989,Potapov1989,Badeeva2003,Kuimov2011,Malysheva2012} and 5) organic addition or substitution reactions \cite{Gusarova1989,Gusarova2001,Trofimov2008,Kuimov2011,Malysheva2012}. 
	However, common photoresist used in lithography process is incompatible with the solution used in Lewis base reactions and organic addition/substitution reactions, and metal contact materials used for BP devices are incompatible high oxidative metal ion reactions. Therefore, only type 1) and type 3) reactions are possible for BP wet etching. 
	
	NaClO and H\textsubscript{2}O\textsubscript{2} were first studied. However, those chemicals etched too fast for few-layer BP, which could completely erode them within seconds as shown in the supplement materials (Figure S1 and S2). Besides, oxygen-induced defects can affect the electronic properties of black phosphorus. % cite the Advanced Functional paper electrochemical method exforliate black phosphorus
	
	% this paragraph is wrong.
	Then, halogen reactions in organic solution were investigated. Considering fluorine, chlorine and bromine are too active with photoresist and incompatible with cleanroom environment, we tested iodine solutions on BP. Considering  that BP degrades when exposed to water and oxygen ambients, an organic solvent was needed \cite{Yang2018}. Therefore, acetone was chosen as the solvent. After etching, 100\% isopropyl alcohol (IPA) was used to rinse the samples. The samples were dried in air. As shown in Figure 1 (a) and (b), clear thinning effect were observed after a BP sample on a silicon substrate had been in a 5 g/L iodine/acetone solution for 5 minutes.  The Raman spectrum in Figure 1 (c) showed good crystallinity of the sample after etching. 
	
	\begin{figure}
		\includegraphics[width=0.22\textwidth]{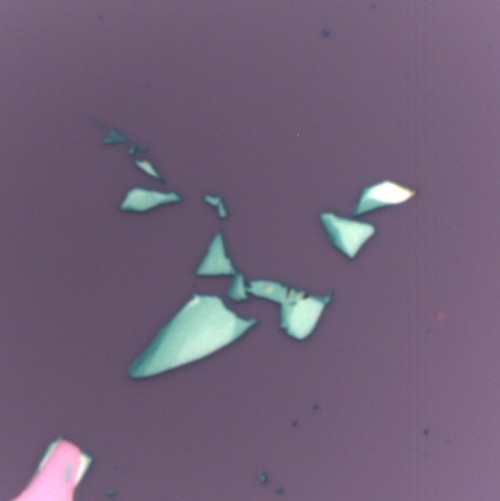}
		\includegraphics[width=0.22\textwidth]{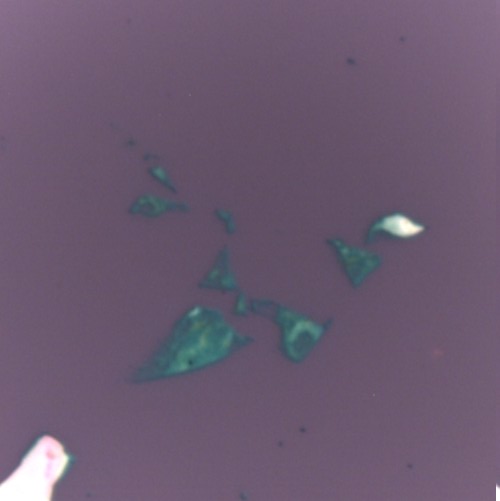}
		\includegraphics[width=0.35\textwidth]{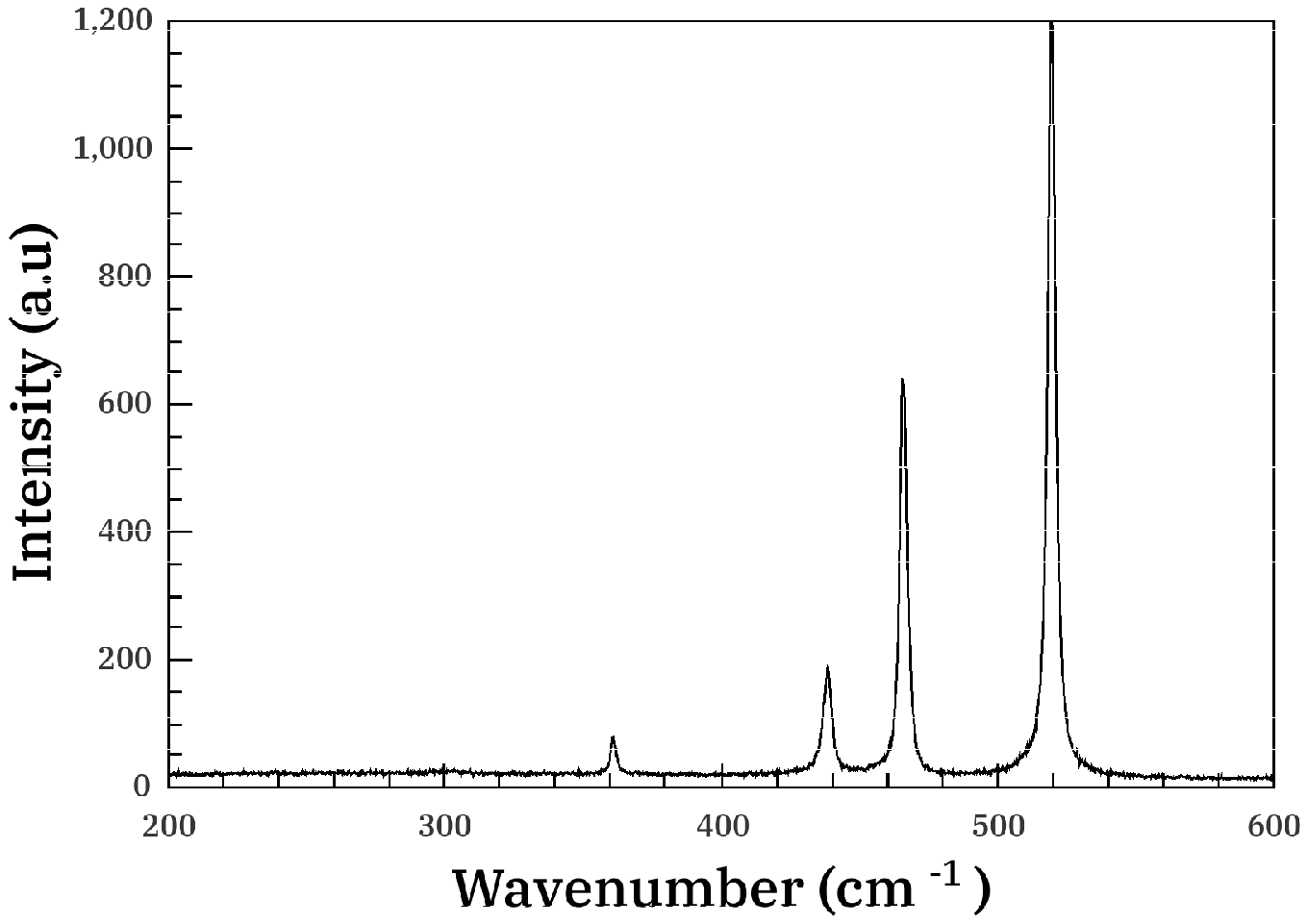}
		\caption {Optical images of samples {a) before thinning, b) after thinning. Color change indicate the etching effect on sample. c) Raman spectra after thinning showing good crystalinity. The etch solution used was 5 g/L iodine/acetone solution.}
		}
	\end{figure}

	However, there were still problems in using acetone as the solvent, such as etching controllability, incompatibility with lithography processes, and the effects on the BP quality, which were investigated and discussed in the next section.

	\begin{figure}
		\includegraphics[width=0.23\textwidth]{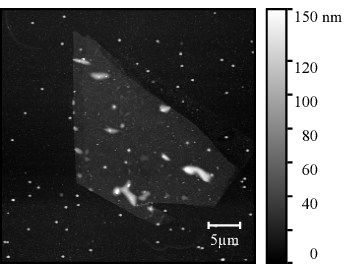}
		\includegraphics[width=0.23\textwidth]{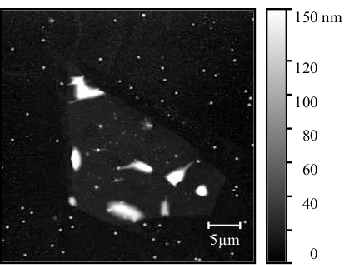}

		\includegraphics[width=0.23\textwidth]{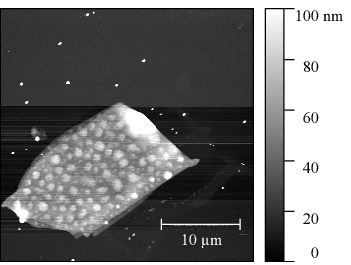}
		\includegraphics[width=0.23\textwidth]{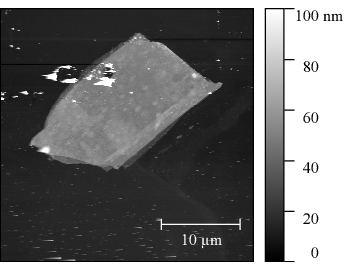}
		\caption{ AFM height images of BP samples. \\
			a) and b): Sample A before and after iodine/chloroform solution etching. Several large bubble-like regions with more than 100 nm height differences are damages formed during the etching. \\
			c) and d): Sample B before and after iodine/IPA-methanol solution etching. The rough surface in c) should be a result of surface oxidation during the sample preparation. Sample B after etching has a very clean and flat surface in d) when compared with those etched by chloroform solutions.}
	\end{figure}
	
	\begin{figure}
		\includegraphics[height=0.12\textheight]{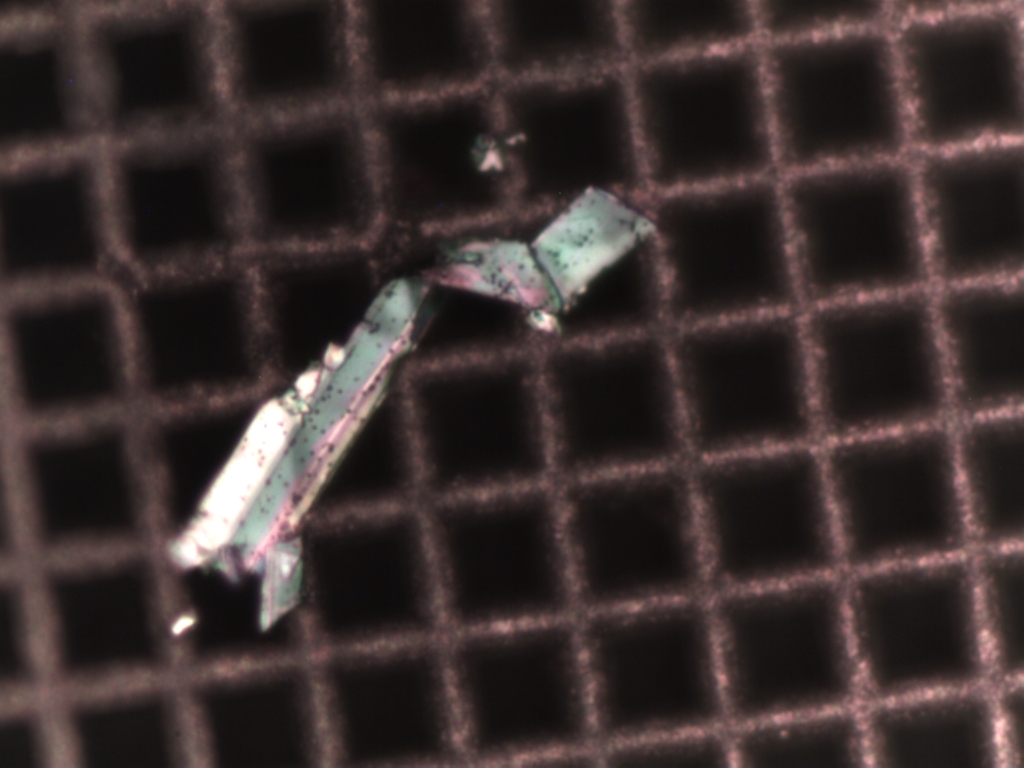}
		\includegraphics[height=0.12\textheight]{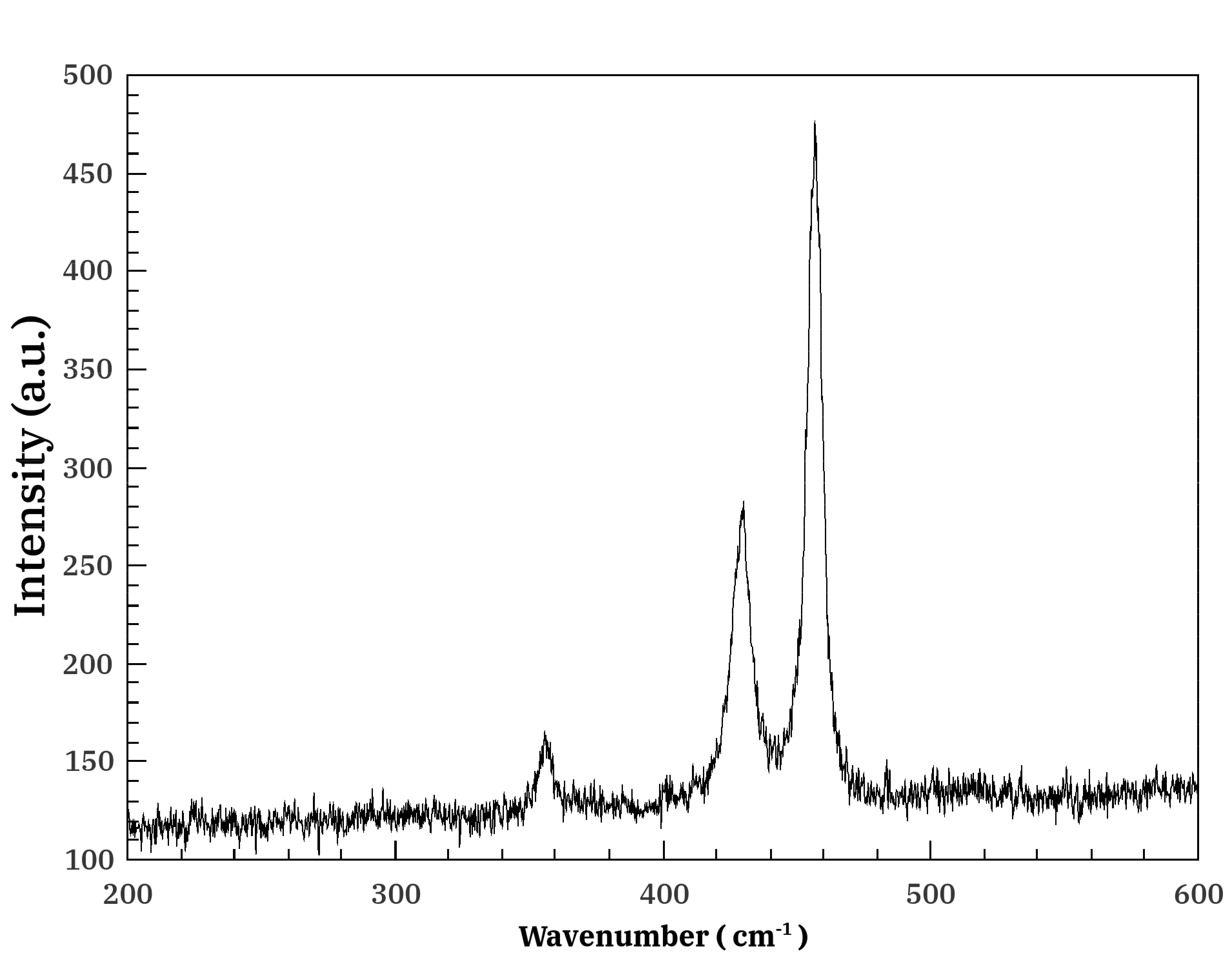}
		\includegraphics[height=0.12\textheight]{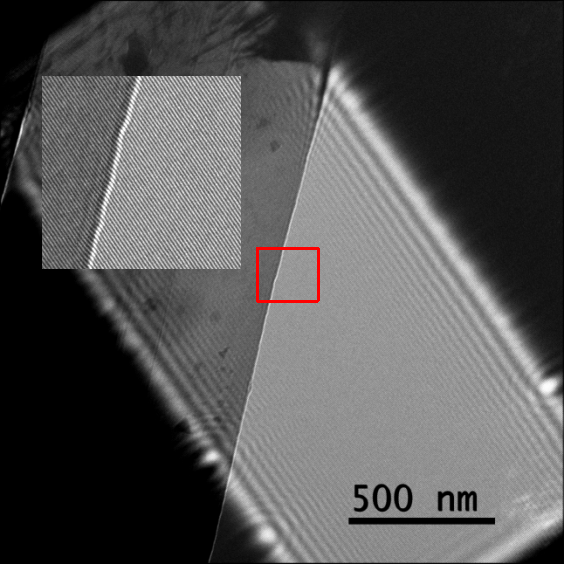}
		\includegraphics[height=0.12\textheight]{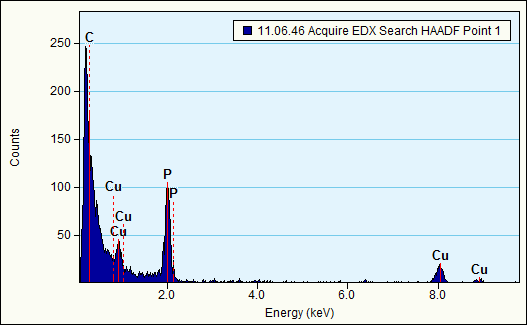}
		\includegraphics[height=0.14\textheight]{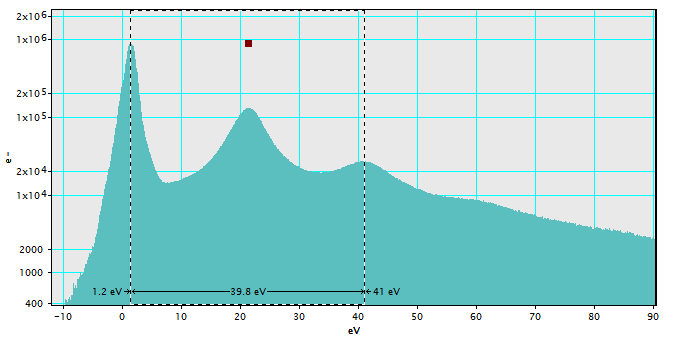}
		\caption{Material characterizations of a BP sample C after etching by a 10 g/L iodine/IPA-methanol solution for 10 mins: a) optical image of the BP sample on a copper grid, b) Raman spectrum and c) STEM image, both showing a good BP crystalinity, d) EDS data showing that no iodine residues left. The carbon peak is from the the polycarbonate (PC) film remain on the copper grid, and e) EELS spectrum.}
	\end{figure}
	\begin{figure}
		\includegraphics[width=0.22\textwidth]{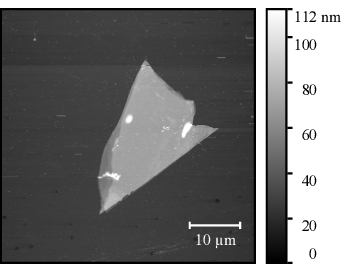}%Aug-30. S5. 20min
		\includegraphics[width=0.22\textwidth]{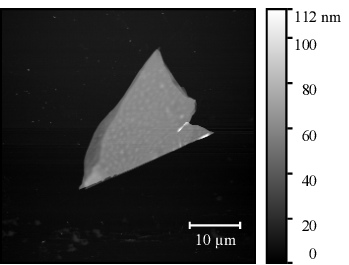}%Aug-31. S5. 20min
		\caption{%\label{fig2} 
			AFM images of Sample D a) before and b) after a 5 g/L iodine/IPA-methanol solution etching of 20 min.  
		}
	\end{figure}
	
	\section{Solvent, concentration, and time parameters for etching}
	\subsection{Solvent dependence}  
	A few nanofabrication-compatible organic solvents  were tested to see their effects on etching process. The reaction time was 10 min and the concentration used for all solutions was 10 g/L.   Acetone, chloroform and IPA-methanol mixture (IPA:methanol=1:1 in volume) solutions were observed to have an apparent thinning effect. 
	
	The iodine/acetone solution etched too fast. It completely etched the 50 nm thick sample away within 10 min. Besides, acetone could remove most kinds of positive photoresists, which makes it incompatible with photolithography process.
	
	The iodine/chloroform solution etched the sample by 10 nm (from 20 to 10 nm in the average thickness) as shown in the Figure 2 (a) and Figure 2 (b). % Add the etch rate from the AFM data ====. 
	However, large bubble-like regions were developed during the etching process. Chloroform was believed to be the reason for BP damage due to its halogen solvent property \cite{Yasaei2015}. 	Similar as acetone, chloroform is incompatible with common photoresists, which makes it unsuitable for photolithography.
	
	The iodine/IPA-methanol solution also etched the sample around 10 nm in 10 mins (from 50 to 40 nm) . The surface was smoother after etching as seen in Figure 2 (c) and (d). The reason was that iodine removed the oxidized BP. The etch rate was suitable for a atomic layer etching (ALE) process. 
	% PLEASE NOTE: the TEM and raman image were different sample!!!!! AND sample in fig 4 were also dfferent from Fig 2, Fig 3.  
	To check the crystallinity and impurities, Raman spectroscopy, STEM, EDS and EELS were performed. Sample C used for STEM imaging, EDS and EELS was fabricated using an modified dry-transfer method as developed in our earlier work.\cite{Fang2018a}
	As seen in Figure 3 (b) and (c), Raman and high resolution STEM image characterizations of Sample C confirmed that the good crystallinity remained after etching.  EDS and EELS data show that Sample C was free of iodine residues after etching and rinsing.

	To study the etch uniformity, a sample, Sample D, was chosen, as it had a clear thickness difference and a clean surface before etching. Uniform etching along the layer-stacking direction $<$010$>$ was confirmed in AFM data of Sample D as shown in Figure 4, indicating a layer-by-layer etching even when the original thickness had more than 20 nm thickness difference. The shape also remained after etching.  
	
	Therefore, an IPA-methanol mixture with a 1:1 volume ratio was found to be the most suitable solvent for BP ALE. In all the following discussions, the solutes used were iodine and the solvents used were all IPA-methanol with this volume ratio.

	\subsection{Time dependence}
	The etching thickness and time relationship under 10 g/L concentration is shown in the Figure 5. The etching thickness has a near linear dependence on time, which means that the etch rate is stable. The etch rate is 0.81 nm/min based on the linear fitting, sustainable for BP thinning control. Detailed AFM data and 1-dimensional (1D) thickness profiles are in supplementary information S3(a)-S3(j).

	\begin{figure}
		\includegraphics[width=0.5\textwidth]{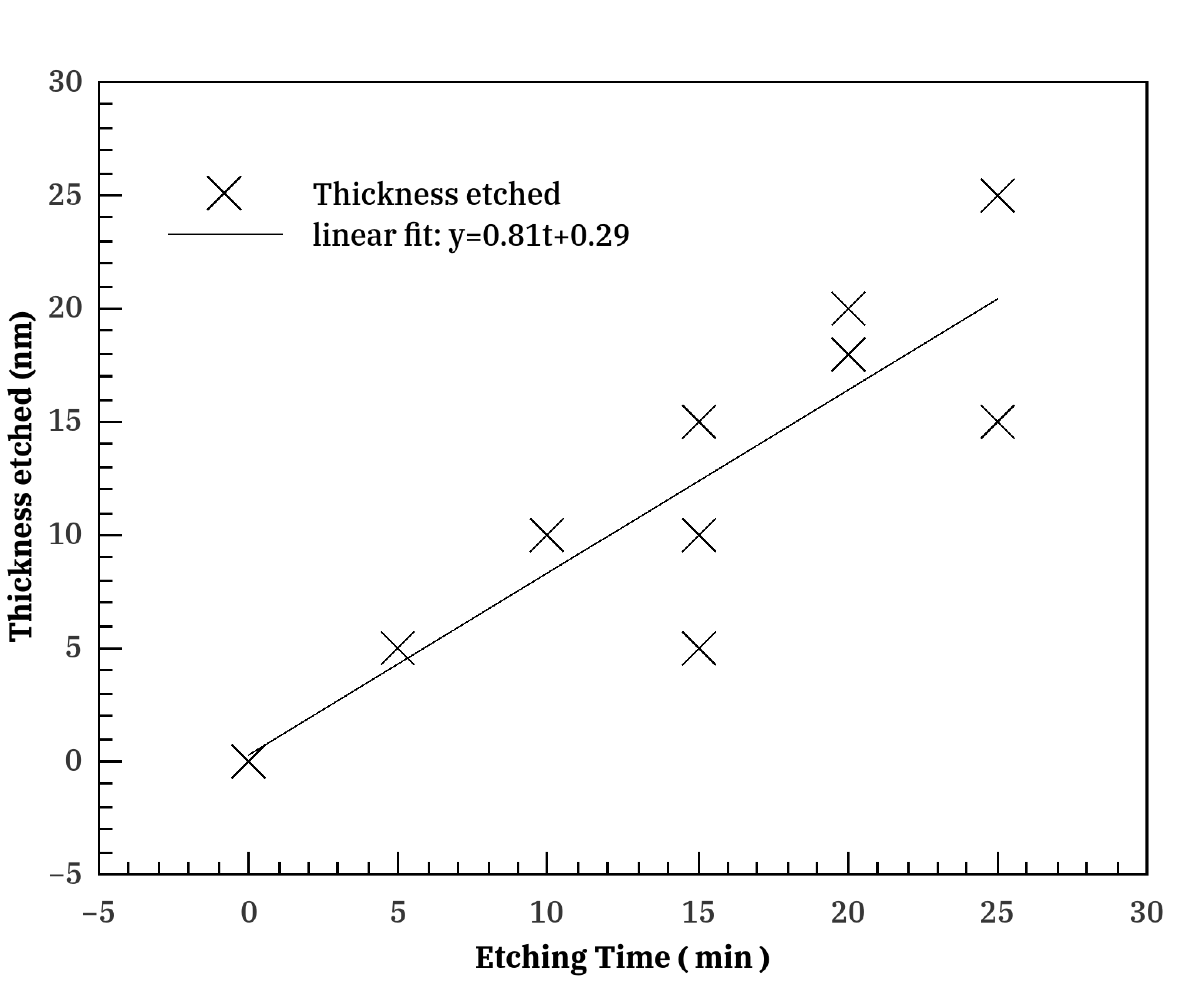}
		\caption{%\label{fig2}
			Thickness etched vs. etch time with a linear fitting. The concentration was 10 g/L. }
	\end{figure}
	\subsection{Concentration dependence}
	With the same etch time of 10 mins, the etching thickness and the iodine concentration relationship is shown in Figure 6. Detailed AFM data and 1-dimensional thickness profilesare shown in supplementary information S4(a)-S4(j). As the etch behavior is near linear with both the etch time and the concentration, we can express this etch behavior as 0.81-1.01 nm/(min*g/L).
	
	\begin{figure}
		\includegraphics[width=0.5\textwidth]{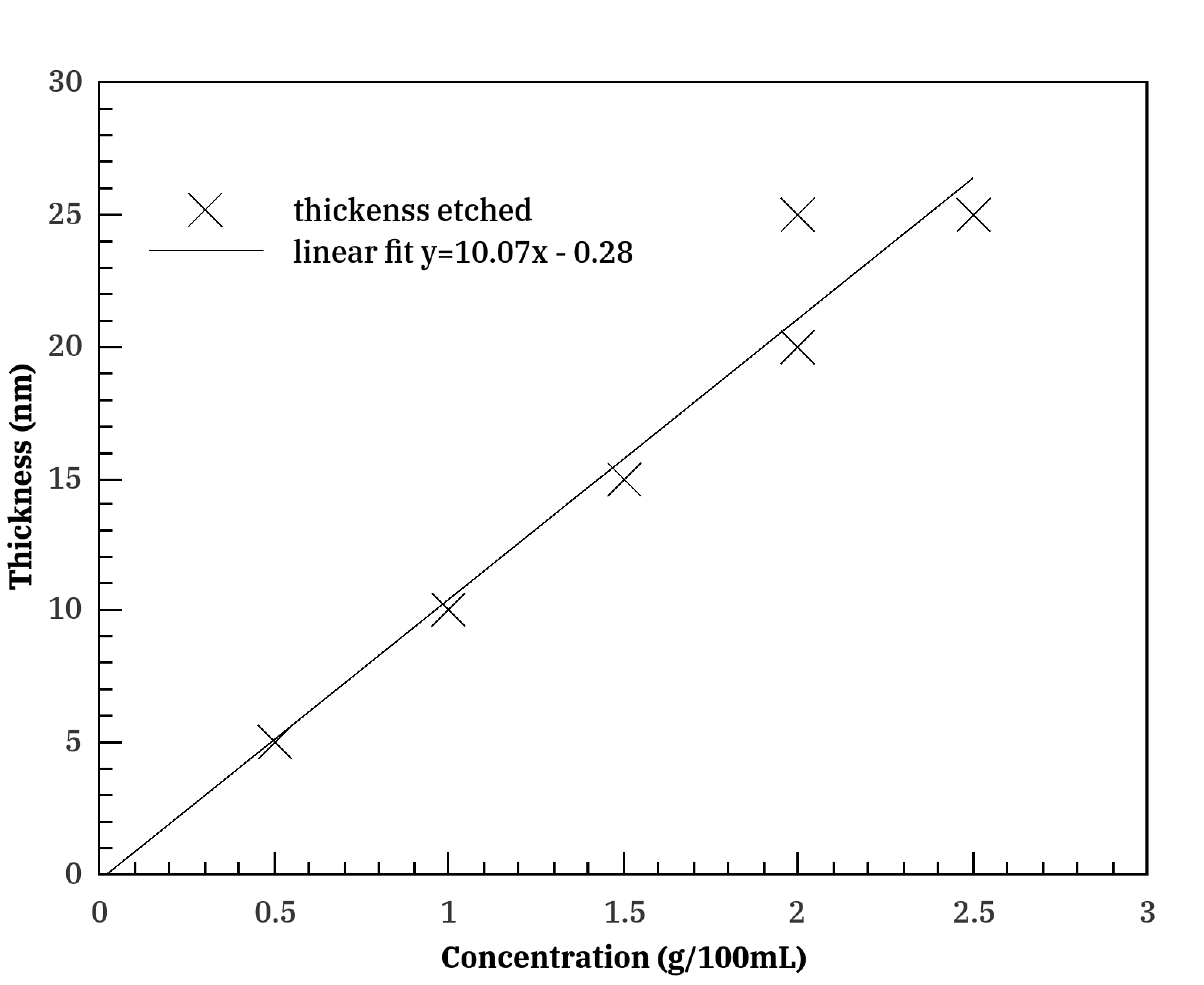}
		\caption{%\label{fig2}
			Thickness etched as a function of the iodine concentration and a linear fitting to the data. The etch time was 10 min.}
	\end{figure}
	
	\section{Lithography and etching results}
	%The aim of the work was to demonstrate an etch process compatible with mainstream photolithography. 
	To obtain a thickness pattern and examine the isotropy/anisotropy of the etching method, photolithography was used. The selection of a suitable photoresist was critical for the ALE process of BP.  Two major problems were encountered during our experiments: 1) IPA and methanol could dissolve most of the i-line and h-line photoresists (AZ-P series photoresists, AZ MIR 703 photoresist, S1800 series photoresists), even after long time hard baking. This greatly limited the photoresist choices; 2) BP can get damaged or oxided at high temperature, which limits the possibility of using hard masks. After some trials, PMMA(polymethyl methacrylate) photoresist was found to be friendly with IPA-methanol etching solutions, which could be used for both DUV lithography and e-beam lithography. 
	
	The lithography and ALE results were shown in Figure 7. The sample was exposed in a low-pressure mercury lamp(253nm DUV light source), developed and putted in 20 g/L iodine solution for 12 min. Then it was taken out from the etch solution when visible color change was observed on the exposed region. According to the AFM line profiles and scanning results before and after etching, we achieved around 20 nm thickness difference, which matches the etch rate we got previously. %=========
	Clear vertical edge could be observed in the result with around 40:1 (nm/$\upmu$m) degree of anisotropy, and pattern shape accuracy is around 0.5 $\upmu$ m. Although planar resolution for the thickness pattern is relatively low comparing with vertical resolution achieved which is atomic layer accuracy, one possible reason is that our DUV lithography condition (mask and light source as discussed in the supplementary material) limited the planar accuracy. Detailed information about our lithography facility and parameter was discussed in supplementary information (S5). 
	%	\begin{figure}
	%		\includegraphics[width=0.2\textwidth]{fig7(with_pr)}
	%		\includegraphics[width=0.2\textwidth]{fig7(etched)}
	%		\includegraphics[width=0.2\textwidth]{fig7(last)}
	%		\caption{a) sample before etching(coated with PR) b) sample after etching(20 g/L iodine solution) with photoresist covered c) sample after etching with photoresist removed}
	
	%\end{figure}

	\begin{figure}
		\includegraphics[width=0.22\textwidth]{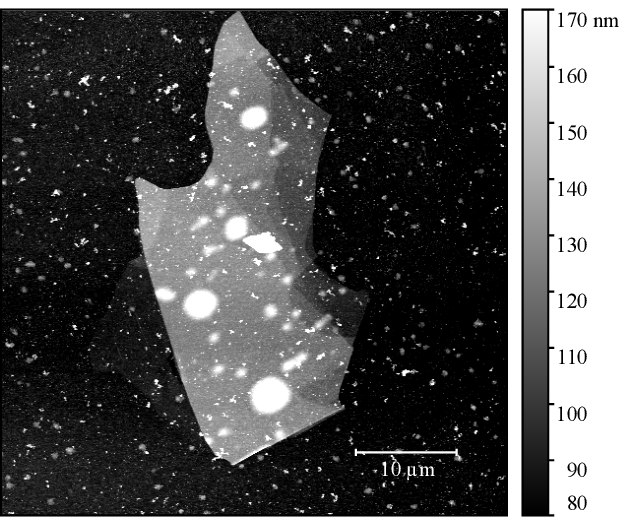}
		\includegraphics[width=0.22\textwidth]{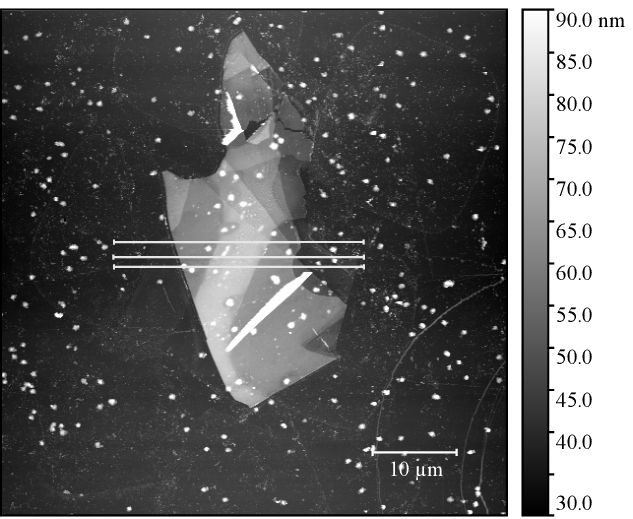}
		\includegraphics[height=0.14\textheight]{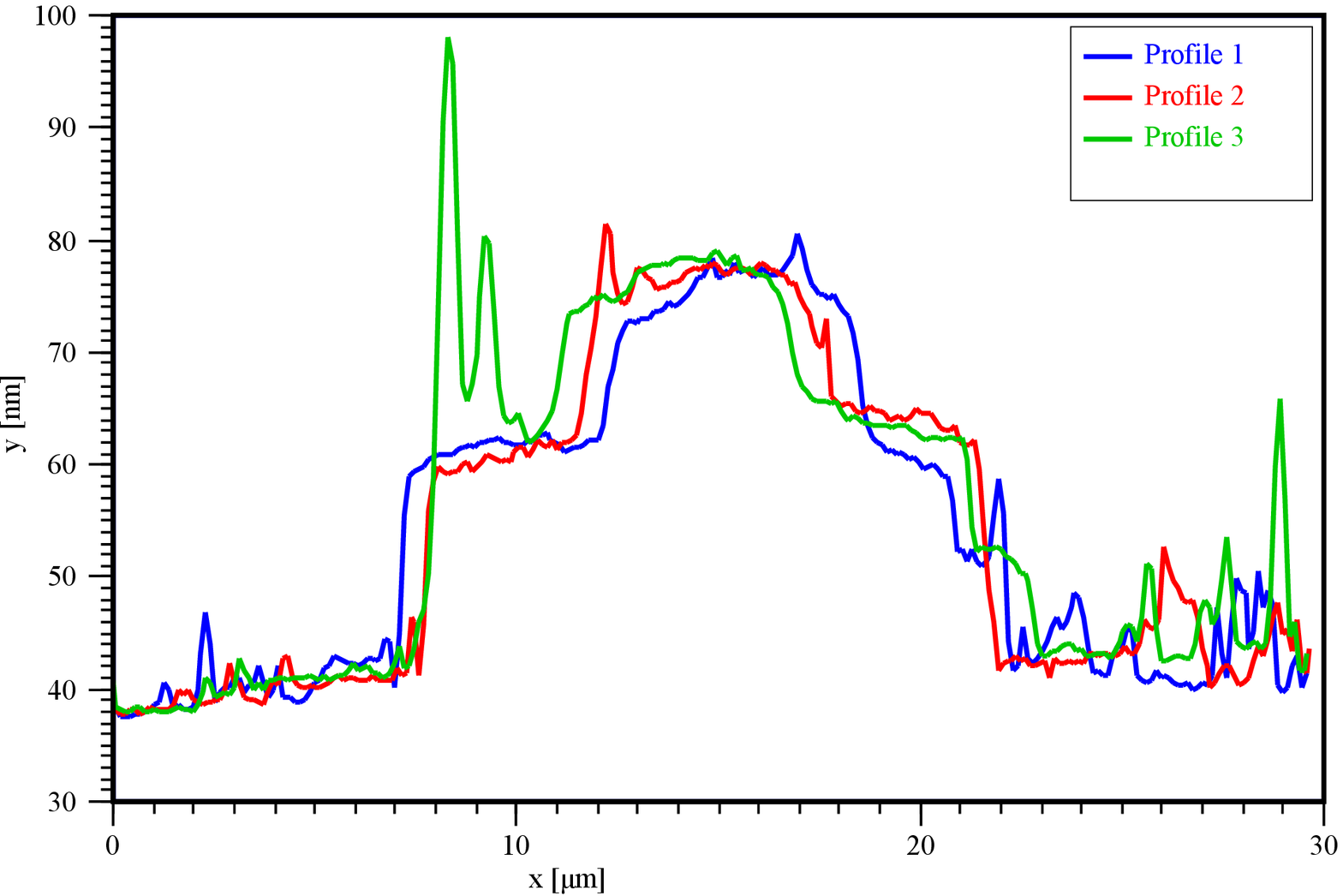}
		\includegraphics[height=0.14\textheight]{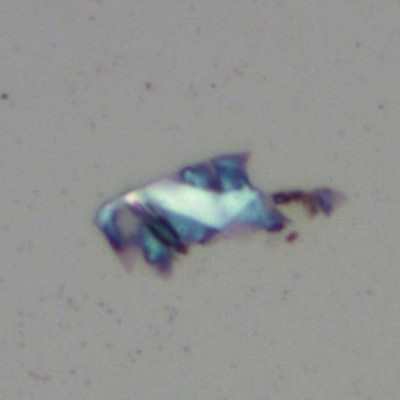}
		\caption{ b) AFM data before etching; d) AFM data after etching (20 g/L iodine solution). The defects on the AFM images could be the photoresist remain on the sample during processing due to the limit of our UVC exposure condition. c) line profile of the thickness, which shown that the edge of our etching shape was within 0.5 um width, very similar to the edge of the sample itself.  e) Optical image of the sample after etching( photoresist removed) }
	\end{figure}
	
	\section{Summary}
	%==== refer to my abstract ====. 
	Atomic layer accuracy BP thickness pattern was achieved using photolithography and  etching method we developed. Organic solution were used as the media and environment for the etching process,  made it possible to get rid of water and oxygen, which are the important sources for the defects in few layer BP. Detailed etching thickness dependence on time, concentration and solution  were studied in this paper, which shown near linear dependence. PMMA was used as the photoresist in this work, which made this method compatible with DUV lithography and e-beam lithography process.  Besides, this method has also shown very reliable thinning ability on BP, which could be used for large area or critically thin few-layer BP fabrication and controlling BP thickness as shown in supplementary Figure S6. 
	
	Therefore, we believe that this BP ALE process provided a low cost, high repeatability way for achieving select region BP layer number engineering which could be used for controlling BP layer number,  fabricating various kind of BP quantum devices, and massively manufacturing large size, few-layer BP samples.
	\section{Experiment Details}
	\subsection{BP sample fabrication}
	BP samples were fabricated by mechanical exfoliation method using Scotch\textsuperscript{\textsuperscript{TM}} Tape. Several very thick samples(120 nm~200 nm thick) for lithography test were first thinned down to around 50 nm thick by our atomic layer etching method to fit our requirement.
	\subsection{Etching solution}
	Etching solutions were prepared in cleanroom condition. To prevent evaporation which would affect the concentration, the etching solution was prepared only minutes before use. Ultrasonic wave were used to shorten the dissolve time.  
	\subsection{STEM sample preparation}
	STEM sample preparation was achieved by two steps. First, we use the dry transfer method designed by our group as we published earlier to transfer the BP sample and make it stiff enough on the grid. Second, we put the copper grid in the etching solution and use tweezers carefully to prevent bending or damage the grid/sample. It should be noticed that this solution also shows etching effect with the copper grid if sample was putted in the solution for more than 30min, while no visible etching effect with gold grid. So if possible we would recommend using gold grid for long time etching process.
	\subsection{Lithography}
	PMMA photoresist was coated on the sample and spinning with 7000rpm rate. We use 1000 mesh TEM grid as the mask for DUV lithography. It should be noticed that hard baking time should be limited as BP could get oxided if it is thinner than 20 nm. Details about our DUV lithography was discussed in our supplement material.
	\section{Acknowledgment} 
	We are very grateful for the kind help and suggestions on lithography from UBC cleanroom and SFU 4D lab, the EELS/EDS characterization data from Professor Kavanagh, and the AFM facility provided from UBC Centre for Flexible Electronics and Textiles and Professor Peymann for some sample's characterization.

	\bibliography{merbib}{}
\end{document}